# Predicting Indian Supreme Court Judgments, Decisions, Or Appeals
## eLegalls Court Decision Predictor (eLegPredict)

*Sugam Sharma, PhD[*], Ritu Shandilya, PhD Candidate[*] and Swadesh Sharma, LLM[#]*
[*]*Iowa State University, USA*, [#]*Additional Prosecution Officer, INDIA*
*Contacts: {sugam.k.sharma, hritussharma, adv.swadesh.sharma}@gmail.com*

**SUMMARY**

Legal predictive models are of enormous interest and value to legal community. The stakeholders, particularly, the judges and attorneys can take the best advantages of these models to predict the case outcomes to further augment their future course of actions, for example speeding up decision making, support arguments, strengthening the defense, etc. However, accurately predicting the legal decisions and case outcomes is an arduous process, which involves several complex steps- finding suitable bulk case documents, data extracting, cleansing and engineering, etc. Additionally, the legal complexity further adds to its intricacies.

In this paper, we introduce our newly developed ML-enabled legal prediction model and its operational prototype, *eLegPredict*; which successfully predicts the Indian supreme court decisions. The *eLegPredict* is trained and tested over 3072 supreme court cases and has achieved 76% accuracy (F1-score). The *eLegPredict* is equipped with a mechanism to aid end users, where as soon as a document with new case description is dropped into a designated directory, the system quickly reads through its content and generates prediction. To our understanding, *eLegPredict* is one of the first few legal prediction models trying to predict Indian supreme court decisions.

**Keywords.** eLegalls, prediction, Indian, supreme, court, judgments, decisions, appeals

## 1. INTRODUCTION

Despite several changes coming through Indian jurisprudence such as e-courts (Verma, 2018), the Indian legal system is not expected to be much different arguably from the legacy system and largely to remain inefficient, complex, and quite slow; where in general, the court proceeding time spans expand to several years to decade. The historical evidences suggest, often for several cases, the court processing literally spans over several decades before reaching to the final verdicts (PTI, 2021). The poor percentage of presiding judges vs. population size is often being monologued, perhaps since India's independence, as the primary basis for unbelievably large period of judicial proceedings and final decision making; however, this argument does not better serve the community in any manner, and society at large continues to suffer due to delayed justice system, which then erodes their constitutional rights, which ensure and safeguard fair and faster justice. In addition to several latent sufferings because of this, the valuable time, energy and resource loss, substantial oneself agony, critical financial stress, which rapidly inches to indebtedness are some prominent human states one undergoes through; and consequently, over the period, the yearnings of civil society for swift justice commence to dry and ultimately die.

Therefore, the Indian legal system has enormous space for enhancement and needs immediate focus and collaborative efforts from the interdependent scientific, more specifically computational and legal, communities to find disruptive methods and mechanisms to advance the civil justice system significantly. Lately, computer and legal scientists together have explored numerous possibilities and opportunities of applying Artificial Intelligence (AI), Machine Learning (ML), Deep Learning (DL), and Natural Language Processing (NLP) techniques and technologies into legal system, which has been mostly remained insulated from Information Technology (IT) world.

A few futuristic scientists have taken up the hard task to apply ML and NLP techniques to develop prediction models to accurately predict court decisions with several eventual goals to advance, improve and elevate the civil justice system; and among others one prominent goal is to significantly shorten the time period of judicial proceedings and judgment delivery. And in this view, devising and development of efficient and accurate legal predictive models become urgent and of utmost importance to better serve the society. In courts, justices and judges can consider and use prediction models as augmented support to make their decision making process faster. The lawyers or attorneys can utilize the prediction models to predict case outcome in advance, which ll help them to better prepare for defense during court arguments. However, given the complex nature of legal system (Ruhl at al., 2017), predicting court judgments and case outcomes with reasonable explanation with greater accuracy is a complex and tedious task and requires intensive general IT, NLP, and ML usage for data - download, extracting, cleansing, modeling, engineering - and model development, training and testing. Therefore, we also endeavor to contribute to this challenging task and develop an efficient prediction model to closely predict Indian supreme court judgments, decisions or appeals. And, in this paper, we discuss our newly developed ML-enabled legal prediction model and its operational prototype, called *eLegPredict*. which successfully predicts the Indian supreme court decisions. The *eLegPredict* is trained and tested over 3072 supreme court cases and has achieved around 76% accuracy. The *eLegPredict* is equipped with an interface (a directory) to support and facilitate end users in accepting their new case description as input; the users can drop their case description in pdf format into that designated directory and as soon as the document with new case description is dropped, the system swiftly reads through its content and generates prediction.

The rest of the paper is structured as follows. Section 2 discusses the related work with special focus on court decision predictions. Section 3 is the architecture of *eLegPredict*. Section 4 discusses the core technical work, the main thrust, of this paper. Section 5 expands on interesting discussion on applied ML and NLP into legal system. And finally, the paper is concluded in section 6, briefly outlining the future work.

## 2. RELATED WORK

Katz and his colleagues (2017) develop a prediction model to predict the voting pattern for SCOTS; Among others, authors also use random forest algorithm with over sixty years SCOTS judgments and claim to achieve 69.7% accuracy on decisions. André Lage-Freitas et al. (2019) research and develop a prediction methodology to forecast the judgements from Brazilian court; the author also code and create an operational prototype, which demonstrates around 79% when trained and tested on 4,043 Brazilian court cases. O'Sullivan and Beel (2019) generate ML-enabled models to predict whether ECHR judgements violate any human rights article; to obtain better performance, the authors use word (echr2vec) and paragraph (doc2vec) embeddings and overall 68.83% test accuracy is achieved. Pillai and Chandran (2020) explore the usage of NLP, with particular emphasis on Bag of Words, and CNN to churn out the important words from Indian court judgments, to classify them – bailable vs. non-bailable - and subsequently to predict their classification; their prediction model achieves around 85% accuracy. Sert et al. (2021) apply AI technique, with NLP's word embedding, to predict whether the judgments from Turkish constitutional court on individual cases violate any public morality and freedom of expression; the authors claim around 90% accuracy in making correct prediction. Sharma, Chudhey, and Singh (2021) apply ML for a very interesting problem, the increasing divorce cases in India; the authors aim to predict the divorce cases, particularly to aid therapists and marriage counselors as the growing divorce number is exerting stress on them; the authors formulate the divorce issue as binary yes/no classification problem and apply their algorithm to predict the divorce possibility with 98.5% accuracy; this algorithm will particularly assist therapists to gauge the gravity of situation between involved parties, and subsequently to strategically plan and guide their counseling efforts in effective direction.

### 3. ARCHITECTURE

Figure 1 depicts the complete architecture of *eLegPredict*. The figure clearly demonstrates the main players, who can take the best advantage of this model- judges, lawyers or attorneys and the common users with some legal knowledge. The content in existing supreme court judgments consist of enough text, which contributes nothing towards the prediction formation; therefore, such text needs to be cleansed. As shown in architecture, the pdf copies of the court judgments are read as input and cleansed under NLP processing. Subsequently, other required operations within NLP realm are also performed to deduce the most efficient vector data, which is to be infused into ML; the intermediary NLP operations are further discussed in greater detail in next section.

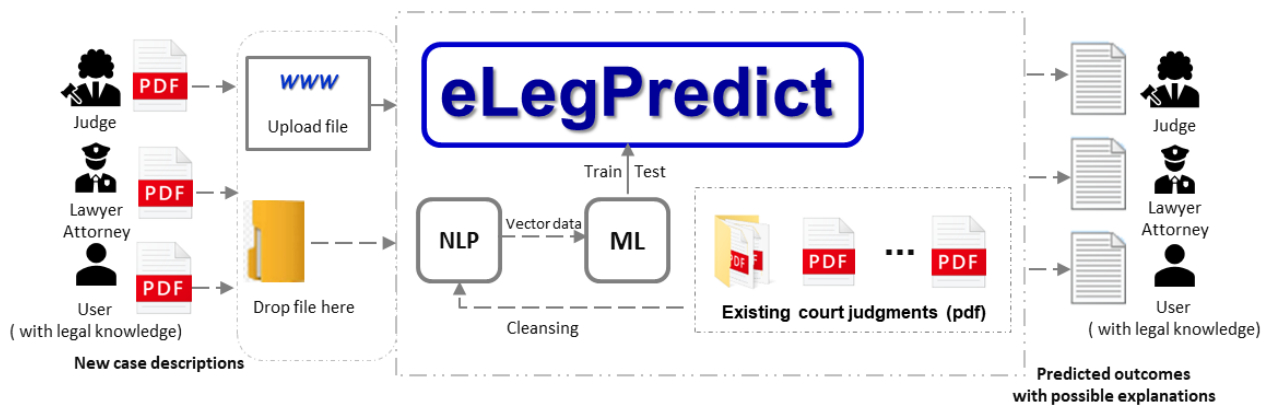

**Figure 1.** *eLegPredict* architecture

In its full avatar, the well trained and tested *eLegPredict* is to provide two interfaces, web and non-web, to the end users for interaction and use. The legal actors can write the new case description in any medium, preferably word document, which could be transformed into a pdf document. The *eLegPredict* accepts the input documents in pdf format only and is to emit the predicted outcomes in easily readable simple text format. In addition to the classification nature of predicted judgment, possible explanation behind the predicted decision is also to be reported in outcome.

### 4. MODEL DESCRIPTION AND PROCEDURE

#### 4.1 Case documents
The very first step in developing any prediction model to find and gather the required data. This work requires the Indian supreme court judgments; therefore, these documents are downloaded (in pdf file format) from *IndianKanoon.org* and a total 3072 cases (final clean documents) are collected in the corpus.

## 4.2 Classification label

The second crucial step is to decide the important attribute(s), which model should predict. Therefore, an extensive survey is carried out through the large corpus of judgments. Based on outcome, the attribute of importance emerged is "appeal", which is mainly classified into three categories -1) allow, 2) dismiss, and 3) dispose. And accordingly, in this work we use this classification to label the court decisions.

## 4.3 NLP modeling and data preparation

This is one of the core tasks of this work, which helps to read through all the 3072 case documents (pdf file format), executes several complex intermediary subroutines and processes and eventually produces clean vector data with appropriate labelling. To obtain better semantic representation of words and to enhance the efficiency, important NLP techniques exploited in this work are outlined here - removal of smaller (less than three characters) words, removal of stop words, removal of white-spaces and punctuation, word stemming, removal of digits, conversion of all words into lower case, which improves word recognition. In sum, the content, which is not important in decision making process is removed. Subsequently, file is tokenized and n-grams (where n= 1 to 4) are generated. For modeling purposes, the textual representation of the words needs to be quantified into a numeric form, therefore, for this the efficient TF-IDF (Term Frequency-inverse Document Frequency) technique is used to quantify the word to capture its relevancy in document and corpus at large. The TF-IDF is tuned to ignore the words with lower frequency (<10%). This helps to further refine data and keeps only the high representative data. As part of the NLP modeling process, we also develop a procedure, which screens through the target sentences of case document and auto captures the decision classification of each document in corpus and appends it to its tokenized vector representation as label. And, in the end, it produces a clean csv data file.

## 4.4 Classifier modeling

Our prediction model, *eLegPredict*, relies on the supervised ML-based classifiers and in this work following leading classifiers are used - eXtreme Gradient Boosting (X Gradient Boost), Neural Network, Support Vector Machine, and Random Forest. The detailed scientific, mathematical and technical discussions on these classifiers are beyond the scope of this paper and readers are advised to look for other online resources.

Each classifier is carefully calibrated with highly optimized hyperparameters. The entire dataset is divided into 80%-20% ratio to train and test the model respectively. The main technical artifacts used in this work are Natural Language Toolkit (NLTK) for NLP, Scikitlearn ML library, and Python. After successfully training the classifiers, mentioned above, a classification report is generated for each one of them and results are reported here.

## 4.5 Results

This section consists of the classification report and normalized confusion matrix of optimized model for each of the classifiers used in this work (Figures 2-5). The results are summarized in decreasing accuracy order. And from the results, it is clear that X Gradient Boost achieves the highest accuracy (F1-score) as 76%, whereas Random Forest demonstrates the lowest accuracy (F1-score).

From the classification report, it is evident that on our dataset, X Gradient Boost classifier performs well as compared to other listed classifiers in this paper. Therefore, we save the trained and tested X Gradient Boost classifier and ignore others. The saved X Gradient Boost classifier aids and facilitates the classification process in *eLegPredict* model to predict the new observations with good accuracy.

**X Gradient Boost classifier:**

|              | precision | recall | f1-score |
|-------------:|----------:|-------:|---------:|
| allow        | 0.77      | 0.82   | 0.80     |
| dismiss      | 0.74      | 0.76   | 0.75     |
| dispose      | 0.75      | 0.61   | 0.67     |
|              |           |        |          |
| **accuracy** |           |        | **0.76** |
| macro avg    | 0.75      | 0.73   | 0.74     |
| weighted avg | 0.76      | 0.76   | 0.75     |

**Neural Network classifier:**

|              | precision | recall | f1-score |
|-------------:|----------:|-------:|---------:|
| allow        | 0.70      | 1.00   | 0.82     |
| dismiss      | 0.00      | 0.00   | 0.00     |
| dispose      | 0.00      | 0.00   | 0.00     |
|              |           |        |          |
| **accuracy** |           |        | **0.70** |
| macro avg    | 0.23      | 0.33   | 0.27     |
| weighted avg | 0.49      | 0.70   | 0.58     |

**Support Vector Machine classifier:**

|              | precision | recall | f1-score |
|-------------:|----------:|-------:|---------:|
| allow        | 0.69      | 1.00   | 0.82     |
| dismiss      | 0.80      | 0.44   | 0.57     |
| dispose      | 0.00      | 0.00   | 0.00     |
|              |           |        |          |
| **accuracy** |           |        | **0.65** |
| macro avg    | 0.50      | 0.48   | 0.46     |
| weighted avg | 0.67      | 0.65   | 0.63     |

**Random Forest classifier:**

|              | precision | recall | f1-score |
|-------------:|----------:|-------:|---------:|
| allow        | 0.58      | 1.00   | 0.73     |
| dismiss      | 0.00      | 0.00   | 0.00     |
| dispose      | 1.00      | 0.25   | 0.40     |
|              |           |        |          |
| **accuracy** |           |        | **0.60** |
| macro avg    | 0.53      | 0.42   | 0.38     |
| weighted avg | 0.52      | 0.60   | 0.48     |

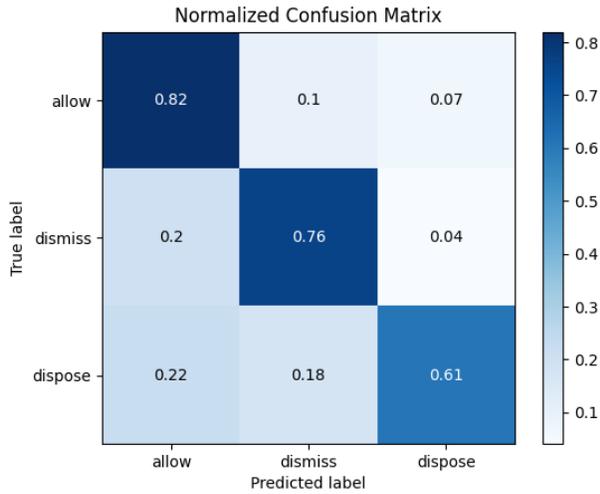

**Figure 2.** X Gradient Boost classifier

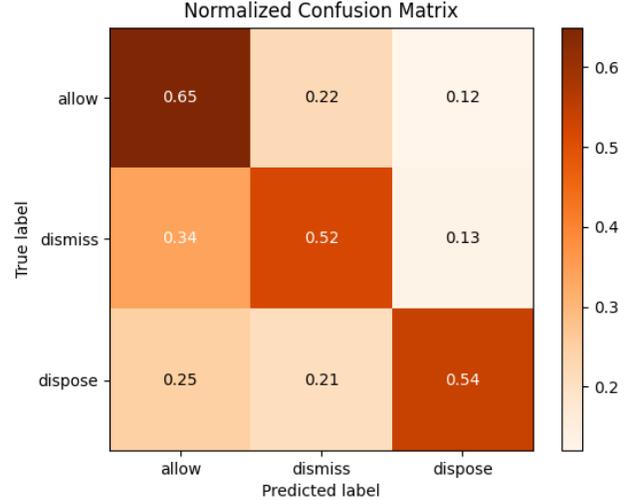

**Figure 3.** Neural Network classifier

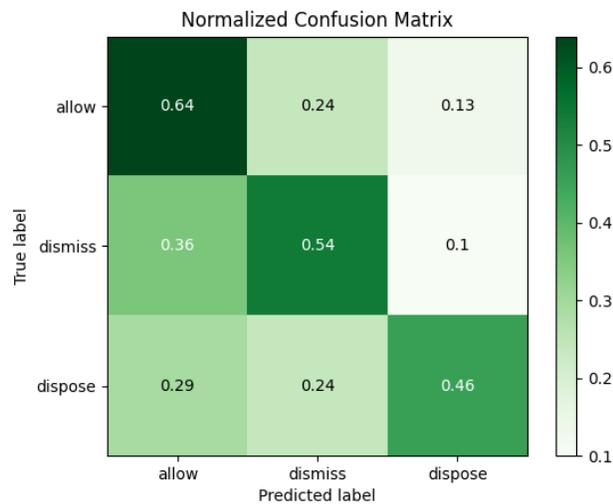

**Figure 4.** Support Vector Machine classifier

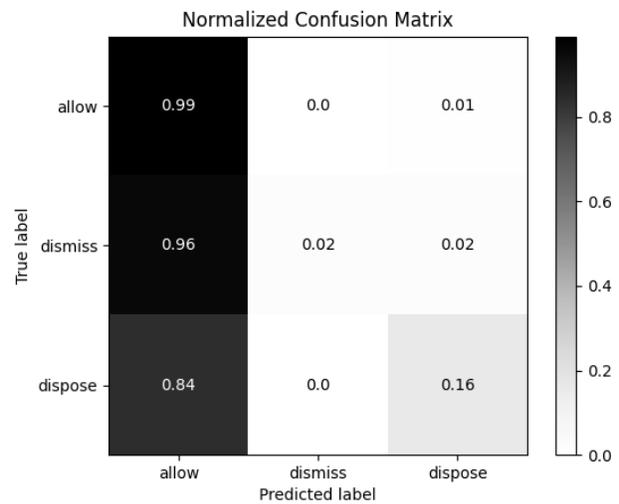

**Figure 5.** Random Forest classifier

At a later stage, with the intent to improve accuracy, we further expanded the dataset and additional supreme court cases from prior years also were downloaded, reaching the total number to 4329 (clean documents). And therefore, we once again repeat the entire process of NLP cleansing and model training and testing. Surprisingly, the performance of all classifiers except X Gradient Boost fluctuates mostly to lower side; however, for X Gradient Boost it almost remains stable (around 76%).

## 5. DISCUSISON

The emerging Legal Informatics (Sharma et al., 2021b)) is penetrating the existing complex legal system to aid tech-enabled efficient processing of legal procedures and resources. The legal system produces enormous amount of textual data through proceedings, opinions, statutes, judgments, legal contracts, etc., which by far has been mostly manually processed, and as a result has been unimaginably sluggish and time consuming. The nature and the format of the legal documents and legal text are quite complex and thus require intensive cleansing process before being used in the LI-world. The legal documents, written in legal language are oftentimes difficult to understand, even for the legal professionals and practitioners, and a user does need good legal knowledge to interpret them in a meaningful way. This is where LI can be effective as legal aid to the general users. In last few years, the LI community has rapidly explored the wider possibilities of applied AI, ML, DL, NLP, etc. into complex legal domain to transform it from traditional manual arena to a technology-equipped modern legal system. And in this transformation phase, the first task, which is underway, is to digitize the existing paper legal documents; given the mammoth number of such documents, this task is expected to continue for decades before the world legal documents become available in digital format. Although, in several countries the digitization procedure has already produced legal datasets and the LI researchers are exploiting it to develop efficient solutions to advance the legal system, for example, LI-enabled prediction models for court judgements, court case classification, legal query-response, lengthy legal document summarization, etc. In the latest progress of LI, there has been substantial discussion on symbol (relationship and events) (Surden, 2018) and embedding (knowledge and facts in vector form) - dependent legal learning (Chalkidis and Kampas, 2019), where the former does it through knowledge interpretation, perhaps suitable for smaller datasets, and the latter exploits hidden crucial features. Also, the embedding-based learning is practically has demonstrated higher performance score though, but not interpretable and thus may introduce ethical concerns and biases in the predicted outcomes; and this sometimes impedes their usage in legal applications. Despite, the symbol and embedding-based learning pose several challenges in LI domain: 1) Legal knowledge interpretation- the content in legal documents is in a structured format and how to extract the interpretable useful legal knowledge from content is always a challenge, 2) Logical reasoning- in LI world, reasoning of an outcome is important to better support and validate prediction. Therefore, it is rather imperative that under LI processing, the existing legal regulations, structures and provisions should remain intact and are not either dismantled or ignored in predictions. As of now, LI research is primarily moving into four directions – court case prediction (Katz et al., 2017; Lage-Freitas et al., 2019; O'Sullivan and Beel, 2019); citation, previous, and similar case search (Xiao et al., 2019; Hong et al., 2020); legal query-response (Kim and Goebel, 2017); catch phrase and text summarization (Mandal et al., 2021; Jain et al., 2021; Bhattacharya et al., 2021). However, there are still enormous challenges in advancing the legal system to a point where it becomes of pervasive usage without any ethical and bias concerns.

## 6. CONCLUSION AND FUTURE WORK

In this research work, we discussed and elaborated on our freshly devised and developed legal prediction model and its prototype, *eLegPredict* to predict Indian supreme court judgments, decisions, or appeals. The *eLegPredict* was trained and tested successfully initially with 3072 (and later 4329) Indian supreme court judgments and attained reasonable accuracy (76%). Also, the prototype was furnished with a non-graphical (directory) interface to ingest inputs, new case documents in pdf format, from end users. So, as user pushed a new case document into interface directory, *eLegPredict* quickly slurped and processed it and proceeded to create prediction about input case.

The future work will be centered around the following. The model has enough scope of accuracy improvement and persistent efforts are being made to achieve it. We also plan to expand our case judgments dataset further and explore other high-performant classifiers with the intent to obtain better accuracy. Additionally, we are working to offer the end users with the reasonable explanation along with the predicted outcome. Also, *eLegPredict* will be equipped with additional web-based graphical interface, which aids the end users to smoothly upload the new case document as input for decision prediction. This *eLegPredict* working prototype will be further developed as an efficient product and is envisioned to be commercialized as an arm of our *eLegalls*.

**REFERENCES**

Bhattacharya, P., Poddar, S., Rudra, K., Ghosh, K., & Ghosh, S. (2021, June). Incorporating domain knowledge for extractive summarization of legal case documents. In *Proceedings of the Eighteenth International Conference on Artificial Intelligence and Law* (pp. 22-31).

Chalkidis, I., & Kampas, D. (2019). Deep learning in law: early adaptation and legal word embeddings trained on large corpora. *Artificial Intelligence and Law*, *27*(2), 171-198.

Surden, H. (2018). Artificial intelligence and law: An overview. *Ga. St. UL Rev.*, *35*, 1305.

Hong, Z., Zhou, Q., Zhang, R., Li, W., & Mo, T. (2020, July). Legal Feature Enhanced Semantic Matching Network for Similar Case Matching. In *2020 International Joint Conference on Neural Networks (IJCNN)* (pp. 1-8). IEEE.

Jain, D., Borah, M. D., & Biswas, A. (2021). Summarization of legal documents: Where are we now and the way forward. *Computer Science Review*, *40*, 100388.

Katz, D. M., Bommarito, M. J., & Blackman, J. (2017). A general approach for predicting the behavior of the Supreme Court of the United States. PloS one, 12(4), e0174698.

Kim, M. Y., & Goebel, R. (2017, June). Two-step cascaded textual entailment for legal bar exam question answering. In *Proceedings of the 16th edition of the International Conference on Articial Intelligence and Law* (pp. 283-290).

Lage-Freitas, A., Allende-Cid, H., Santana, O., & de Oliveira-Lage, L. (2019). Predicting Brazilian court decisions. *arXiv preprint arXiv:1905.10348*.

Mandal, A., Ghosh, K., Ghosh, S., & Mandal, S. (2021). A sequence labeling model for catchphrase identification from legal case documents. *Artificial Intelligence and Law*,

O'Sullivan, C., & Beel, J. (2019). Predicting the outcome of judicial decisions made by the european court of human rights. *arXiv preprint arXiv:1912.10819*.

Ruhl, J. B., Katz, D. M., & Bommarito, M. J. (2017). Harnessing legal complexity. *Science, 355(6332), 1377-1378.*

Pillai, V. G., & Chandran, L. R. (2020, August). Verdict Prediction for Indian Courts Using Bag of Words and Convolutional Neural Network. In *2020 Third International Conference on Smart Systems and Inventive Technology (ICSSIT)* (pp. 676-683). IEEE.

Press Trust of India (PTI), (2021). "Man, 108, Dies Just Before Supreme Court Hears Case He Filed In 1968." NDTV. [Online] *https://www.ndtv.com/india-news/man-108-dies-just-before-supreme-court-admits-case-he-pursued-since-1968-2491789* (access Sept 25, 2021)

Sert, M. F., Yıldırım, E., & Haşlak, İ. (2021). Using Artificial Intelligence to Predict Decisions of the Turkish Constitutional Court. *Social Science Computer Review*, 08944393211010398.

Sharma, A., Chudhey, A. S., & Singh, M. (2021, March). Divorce case prediction using Machine learning algorithms. In *2021 International Conference on Artificial Intelligence and Smart Systems (ICAIS)* (pp. 214-219). IEEE.

Sharma, S., & AL, R. S. (2021a). eLegalls: Enriching a Legal Justice System in the Emerging Legal Informatics and Legal Tech Era. *International Journal of Legal Information*, 49(1), 16-31.

Sharma, S., Gamoura, S., Prasad, M.D., and Aneja, A. (May 25, 2021b). Emerging Legal Informatics towards Legal Innovation: Current status and future challenges and opportunities. Preprint of accepted paper in *Legal Information Management Journal, Cambridge University Press.*

Verma, K. (2018). e-courts project: A giant leap by indian judiciary. *Journal of Open Access to Law, 6(1).*

Xiao, C., Zhong, H., Guo, Z., Tu, C., Liu, Z., Sun, M., ... & Xu, J. (2019). Cail2019-scm: A dataset of similar case matching in legal domain. *arXiv preprint arXiv:1911.08962.*

**AUTHORS AND CONTRIBUTION**

**Sugam Sharma** obtained his PhD in Computer Science from Iowa State University, USA in 2013. His research interests include data science & databases, social & legal informatics and legal tech. Dr Sharma also holds MS in Computer Science from Jackson State University, Mississippi, USA and BE in Computer Science from Roorkee, India. Dr. Sharma has around 14 years of research experience in computing and have published good number of papers in reputed journals. Since 2014, he is actively doing research in social informatics and sustainability and some of the published work, eFeed-Hungers, has received international news and media attention. Dr. Sharma is the founder/CEO of eFeed-Hungers.com. Dr. Sharma's current research is mainly focused on the emerging legal informatics and legal tech and has conceived a research-led legal tech startup eLegalls.

Recently, Dr. Sharma has completed his Executive Education in Legal Tech Essentials from Bucerius Law School, Germany and MCLE in Future Law from Stanford Law School, USA.

In this work, Dr. Sharma has developed the concept of *eLegPredict* and has written code and also has developed this document.

**Ritu Shandilya** is a PhD candidate in computer science at Iowa State University. Her research mainly focuses on Machine Learning, Deep Learning, NLP, etc. She has extensive teaching experience in India and US and instructs Artificial Intelligence and Machine Learning to undergraduate and graduate classes for several years now. Ritu also holds BSc (computer applications), Master of Computer Application (MCA) and MTech (computer science) degrees from India.

In this work, Ritu has provided the support and advice from Machine Learning aspect wherever needed and also has reviewed this document.

**Swadesh Sharma** has completed his legal education, LLB and LLM (Gold Medalist), from India and currently works as Additional Prosecution Officer, Govt. of Uttar Pradesh, INDIA. In the past, Advocate Swadesh Sharma has also been a guest faculty of law at Gautam Buddha University, India.

In this work, Advocate Swadesh Sharma has provided guidance and support in obtaining Indian supreme court case judgments and documents.